\documentclass[prl,twocolumn]{revtex4}

\usepackage{graphicx}
\usepackage{wasysym}

\begin{document}

\title{Proposal for a Topological Plasmon Spin Rectifier}

\author{Ian Appelbaum}
\altaffiliation{appelbaum@physics.umd.edu}
\author{H.D. Drew}
\author{M.S. Fuhrer}
\affiliation{Center for Nanophysics and Advanced Materials and Department of Physics, University of Maryland, College Park MD 20742 USA}

\begin{abstract}
We propose a device in which the spin-polarized AC plasmon mode in the surface state of a topological insulator nanostructure induces a static spin accumulation in a resonant, normal metal structure coupled to it. Using a finite-difference time-domain model, we simulate this spin-pump mechanism with drift, diffusion, relaxation, and precession in a magnetic field.  This optically-driven system can serve as a DC ``spin battery'' for spintronic devices.   
\end{abstract}

\maketitle

The ``topological insulator'' (TI) is a class of strongly spin-orbit-coupled materials with topologically-protected, chiral surface states (of opposite group velocity and spin) crossing a bulk electronic gap.\cite{FUKANE} In these surface states, spin and momentum are perfectly related, so that the momentum asymmetry caused by an induced charge current necessarily spin-polarizes the system.

Recently, Raghu \emph{et al}\cite{RAGHU} have theoretically investigated the properties of resonant electromagnetic excitations of charge carriers (plasmons) in the surface states of these materials. Because of the large oscillating current induced at resonance by coupling to incident periodic electric fields, a potentially large time-dependent spin density is created. It was suggested that these spin-polarized carriers could be used to generate spin currents through nonmagnetic materials in unconventional ``spintronic'' devices, but it is not clear how this could be done since the time-average of the instantaneous polarization is zero. 

In this Letter, we propose a means to rectify the plasmon-induced AC spin accumulation in TIs by spatially segregating them in an adjacent resonant plasmonic nanostructure. By simulating time-dependent spin drift, diffusion, and relaxation, we show that non-equilibrium spin polarizations created by the plasmon in the TI can be preserved in steady-state at the edges of the nonmagnetic metal structure, creating a ``spin battery'' to potentially drive spin-electronic devices.

A schematic illustration of the coupled TI-NM plasmon spin rectifier is shown in Fig. 1. Because the perpendicular component of current must be zero at any boundary, the maximum current (and hence maximum spin density) is at the center of the TI structure during plasmon excitation. At this location, a second plasmonic nanostructure with identical resonance, made from a ``trivial'' nonmagnetic (NM) conductor makes contact. Spin generated in the TI can thus flow via diffusion across the interface into the NM. When spin-up ($S_y>0$) is generated by a positive electron current in the TI, it diffuses into the NM and is carried by drift in the +x direction. Half an optical cycle later, spin down is generated by a negative electron current in the TI and carried by drift in the -x direction in the NM. In steady-state, this spatial segregation will be maintained in the presence of constant spin diffusion and relaxation. 

\begin{figure}
\includegraphics[width=7.5cm, height=3cm]{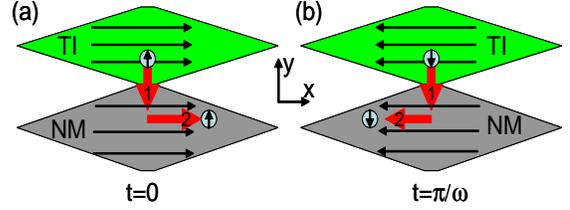}
\caption{  \label{FIG1}
Spin rectification scheme. (a) During one half plasmon cycle, electrons in the chiral topological surface state move in the +x direction and are necessarily in the $S_y$-up spin state. They first diffuse across the interface with a plasmon-resonant normal metal (1) and then are carried by drift in the +x direction (2). (b) During the other half of the plasmon cycle, opposite spin is carried in the opposite direction, thus spatially rectifying the time-varying spin density in the topological plasmon.  
}
\end{figure}

To substantiate this qualitative description, we numerically simulate the spin-rectification mechanism by iterating the one-dimensional spin drift-diffusion-relaxation equation in the NM

\begin{equation}
\frac{dS_y}{dt}=D\frac{d^2S_y}{dx^2}-\frac{d}{dx}\left[v(x,t)S_y\right]-S_y/\tau_s
\label{DRIFTDIFFEQ}
\end{equation}

\noindent with a Crank-Nicolson technique until steady-state is reached. Here, $v(x,t)=v_d \sin(\pi x/L)\cos(\omega t)$, where $v_d$ is the maximum drift velocity of carriers in the NM, $L$ is the length of the nanostructure, and $\omega$ is the plasmon radial frequency. $D$ is the spin diffusion coefficient and $\tau_s$ is the spin relaxation time. As shown in Fig. 2, zero spin-current is maintained at the boundaries of the one-dimensional lattice both by the explicit form of this drift velocity function ($v(0,t)=v(L,t)=0$) but also by mirroring the spin density across the boundary to eliminate spin diffusion. Spin injection into the center of the NM at $x=L/2$ due to diffusion from the TI is modeled by a source proportional to $S_y^{TI} cos(\omega t)-S_y(L/2)$, where $S_y^{TI}$ is the maximum spin density in the TI surface state. All parameters for the NM are taken from known values for Al at room-temperature: D=40cm$^2$/s, $\tau_s$=65ps.\cite{RT_SPIN}

\begin{figure}
\includegraphics[width=7.5cm, height=2.25cm]{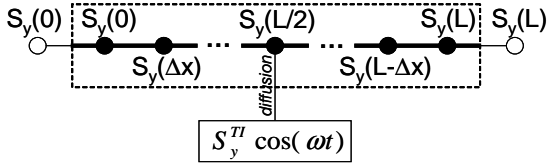}
\caption{  \label{FIG2}
One-dimensional simulation of spin rectification in the NM includes finite-differences Crank-Nicolson solution of the spin drift-diffusion-relaxation equation [Eq. (\ref{DRIFTDIFFEQ})] subject to zero spin-current boundary conditions and diffusive injection of spin from the TI plasmon into the center of the NM plasmon.
}
\end{figure}

We now estimate the plasmon characteristics, treating the surface state of the TI dot as an oblate spheroidal plasmonic element of diameter $L$ and effective thickness $t$ along $\hat{z}$.  The electric polarization of the plasmonic element can be expressed as\cite{LANDAU}

\begin{equation}
4\pi P_x=\frac{(\epsilon-\bar{\epsilon})E_0^x}{1+(\epsilon-\bar{\epsilon})n^x}=4\pi n_2ex_0/t, 
\label{PLASMONPOL}
\end{equation}

\noindent where $E_0^x$ is the incident field, $n^x$ is the depolarization factor of the spheroid for $E_0^x$, $x_0$ is the spatial charge carrier displacement and $n_2$ is the 2-dimensional density of the surface state. The effective static dielectric external to the surface state $\bar{\epsilon}$ we take as the average between vacuum and the bulk TI $\approx$50. The effective dielectric of the surface state $\epsilon$ is taken as $\bar{\epsilon}-\frac{\omega_p^2}{\omega(\omega+1/\tau_k)}$, where $\omega_p^2=\frac{e^2E_F}{\hbar^2t}$. For $t<<L$, $n^x\approx\frac{\pi t}{4L}$.\cite{LANDAU}

The electric polarization is maximum when the denominator of Eq. (\ref{PLASMONPOL}) is minimized, which gives the plasma resonance frequency $\omega^2=\frac{\pi e^2E_F}{4L\bar{\epsilon}\hbar^2}$.\cite{RAGHU} At resonance, $x_0$ is then given by 

\begin{equation}
\frac{x_0}{L}=\frac{E_0^x}{4\pi n_2e}\omega\tau_k,
\end{equation}

\noindent where $\tau_k$ is the momentum scattering time; we assume $\omega\tau_k<1$. The maximum drift velocity is simply $v_d=\omega x_0$. 


For $L$=1 $\mu$m and $E_F$=100 meV, the resonant frequency $f=\omega/2\pi\approx$700 GHz. A relative amplitude of $x_0/L = 0.1$ requires an incident peak power of 100 W focused to 1 mm, which is well within the range of pulsed THz radiation sources. Under these conditions, $v_d$ is on the order of $10^7$cm/s; we assume the modes in the TI and NM are electromagnetically equivalent and use this value in the following simulations.

In Fig. 3(a) and (b) we show the simulation results of $S_y(x)$ during one complete plasmon cycle using the parameters given above after approximately 2.5ns simulation time. These simulations use a spatial discretization $\Delta x=$ 10nm and time step of $\Delta t=$ 10 fs, such that the dimensionless parameters $D \Delta t/ \Delta x^2 ,v_d\Delta t/\Delta x<1$, which is needed for numerical stability. Although the spin diffusion into the center of the NM plasmonic structure is harmonically oscillating in time, the spatial symmetry breaking provided by the plasmon-induced drift gives rise to a static spin accumulation which diffuses to the edges of the NM. The cycle-averaged spin density clearly shows opposite values of $S_y$ on either side, constituting a polarization splitting and spin-battery ``potential'' $\Delta S_y$=0.231 relative to the maximum spin density in the TI, $S_y^{TI}$.    

\begin{figure}
\includegraphics[width=6.5cm, height=10cm]{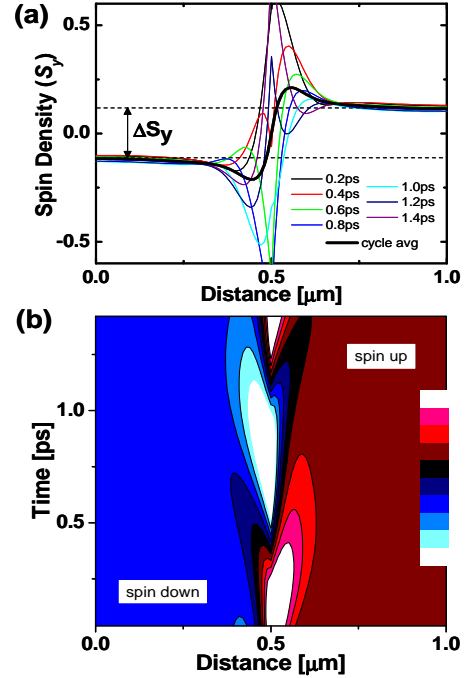}
\caption{  \label{FIG3}
Spatial spin density evolution in a 1-d model of a normal-metal plasmonic structure coupled to a TI in resonance at 700GHz. (a) shows the cycle-average spin density and at several times throughout the cycle with period $T_p\approx 1.4$ps; (b) shows the full evolution with time on the vertical axis. Throughout a full cycle of plasmon oscillation, a steady-state spin accumulation of equal and opposite sign is apparent, thereby constituting DC spin rectification of the AC spin generation from the topological plasmon.   
}
\end{figure}
 
It is useful to determine the effect of changing parameters on this polarization splitting. In Figs. 4(a)-(e), we plot $\Delta S_y$ in black as a function of $v_d$, $\tau$, $L$, $D_n$, and $T_p=2\pi/\omega$ around the fixed values given above. Clearly, to maximize the spin battery output one desires intense electromagnetic illumination causing large drift velocities, a large spin lifetime in the NM, and low plasmon frequencies. 

\begin{figure}
\includegraphics[width=8.5cm, height=11cm]{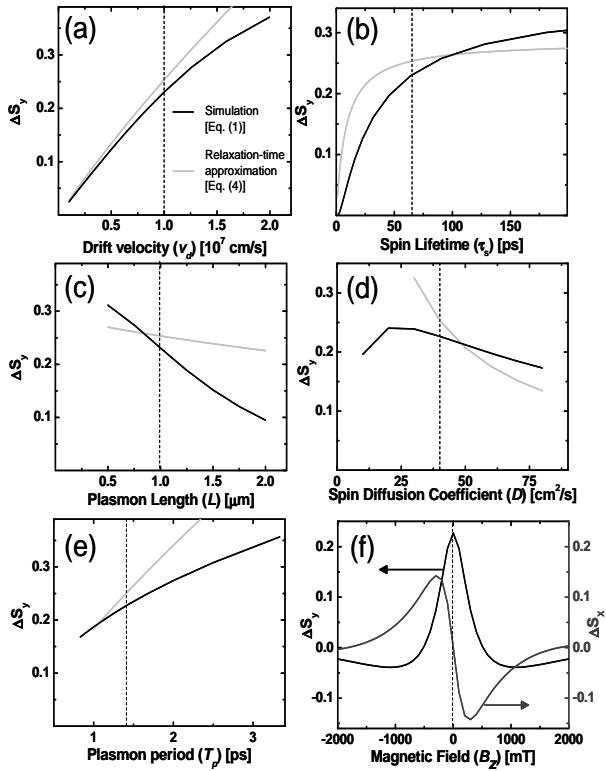}
\caption{  \label{FIG4}
Dependence of spin accumulation $\Delta S_y$ on transport parameters in the NM plasmon nanostructure: (a) drift velocity $v_d$, (b) spin lifetime $\tau_s$, (c) length $L$, (d) Diffusion coefficient $D$, and (e) plasmon period $T_p$. Black lines are the results of Crank-Nicolson simulations of Eq. (\ref{DRIFTDIFFEQ}) and grey lines are the heuristic predictions of Eq. (\ref{RELAXTIMEEQ}).  Vertical dashed lines indicate values used in calculations shown in Fig. 3. In (f), a perpendicular magnetic field $B_z$ causes spin precession and dephasing.
}
\end{figure}

The relaxation-time approximation can give a heuristic prediction of all the trends shown in Figs. 4(a)-(e). Spin-up and spin-down regions are separated by periodic carrier drift by a distance of $\approx v_dT_p$. The loss due to (\emph{i}) spin relaxation in the region towards the boundary at a rate $\approx \frac{\Delta S_y}{2 \tau_s} (L/2-v_dT_p/4)$ and (\emph{ii}) spin diffusion from spin-up to spin-down regions equal to $D \frac{\Delta S_y}{v_d T_p}$ must be balanced in steady state by spin injection. The latter proceeds by diffusion and hence is proportional to $(S_y^{TI}-\Delta S_y/2)$. Therefore, in the weak injection limit, we expect that the spin-battery strength scales as  

\begin{equation}
\Delta S_y\propto \left[\frac{D}{v_dT_p}+\frac{1}{2\tau_s}\left(\frac{L}{2}-\frac{v_dT_p}{4}\right)\right]^{-1}.
\label{RELAXTIMEEQ}
\end{equation}

\noindent Favorable comparison of the trends predicted with this method is made to the numerical model in Figs. 4(a)-(e) (grey lines). 

Our simple spin drift-diffusion-relaxation model suggests some caveats. For instance:
\begin{enumerate}
\item{The validity of a one dimensional approximation requires the diffusion time across the interface should be faster than half the plasmon cycle. For the parameters used here (D=40 cm$^2$/s and f=700GHz), the allowed length $\sqrt{D/f}$= 38nm. This will require nanostructures of high aspect ratio. The detailed consequences of lower aspect ratios can be simulated with a two-dimensional extension of the spin drift-diffusion model described above.}

\item{The spatial overlap between TI and NM nanostructures in our simulations is modeled via spin diffusion into a single finite-difference element. In practice, the spatial overlap should be smaller than the carrier drift distance $v_d/f$ (here 143nm, well within the range of electron-beam lithography).}

\item{Matching the plasmon resonance frequencies between the TI and NM structures is crucial, and is aided by (anti-)symmetric mode coupling and phase-locking.  (For small structures where $L<<c/\omega$, only the symmetric mode is excited by the incident field.) In practice, the TI resonance may be tuned by electrostatic gating if the gate itself does not prohibitively interfere with the plasmon mode. This gate can also be used to adjust $E_F$ below the Dirac point where there will be a second resonance (of spin-polarized holes).}

\item{The effect described here is dependent on the strength of the oscillating spin density $S_y^{TI}$ in the TI driving diffusion into the NM. The spin polarization $P\approx\Delta E/E_F= eE_0^x\lambda/E_F=v_d/v_F$ (where $\Delta E$ is the Fermi-level asymmetry between spin-up and -down states, $\lambda$ is the mean-free-path, and $v_F$ is the Fermi velocity) is $\approx$ 0.1 here. Significant increases in this value may be limited by energy relaxation through strong optical phonon coupling; for instance, in Bi$_2$Se$_3$ at $\Delta E\approx$8 meV.\cite{BI2SE3PHONON}}
\end{enumerate}

To experimentally confirm the presence of the expected spin accumulation in the NM, we propose to detect nonzero $\Delta S_y$ with ferromagnetic tunnel voltage probes.\cite{JOHNSON85} To rule out spurious signals due to asymmetric tunnel barriers and current rectification/photovoltaic effects across the contacts, a magnetic field $B_z$ perpendicular to the surface (in the $\hat{z}$ direction) will be used to induce spin precession and subsequent dephasing (Hanle effect), suppressing this voltage signal.\cite{JOHNSON85} By modeling evolution of both $S_y$ and $S_x$ spin components, and by incorporating spin precession in Eq. (1) by adding a term $-\frac{g\mu_BB_z}{\hbar} \hat{z}\times (S_x\hat{x}+S_y\hat{y})$,\cite{SPIN_CN} (where $g=2$ is the electron spin g-factor, $\mu_B$ is the Bohr magneton, and $\hbar$ is the reduced Planck constant) we have simulated this suppression, shown in Fig. 4(f) for both $\Delta S_y$ and the spin signal in a perpendicular, in-plane direction $\Delta S_x$. In general, longer spin lifetimes will result in stronger low-B-field suppression. 

Finally, we wish to point out that this plasmon-induced spin generation effect is not necessarily confined to TIs. For example, the spin-Hall effect has been observed in extrinsically doped n-GaAs.\cite{KATO} Excitation of plasmons in GaAs will cause a transverse AC spin accumulation; subsequent diffusion of these spins into resonant nanostructures with limited spin-orbit interaction strength is expected to result in spatial rectification as in the above scenario. 

We acknowledge the support of the NSF-MRSEC at the University of Maryland.

\end{document}